\documentstyle[11pt,epsfig,amsfonts]{article}
\title{Self force in 2+1 electrodynamics}
\author{Yurij Yaremko\footnote{Electronic mail: yar@ph.icmp.lviv.ua}}

\date{Institute for Condensed Matter Physics, National Academy of 
Sciences of Ukraine, 1~Svientsitskii St., 79011 Lviv, Ukraine}

\begin{document}

\maketitle

\begin{abstract}
The radiation reaction problem for an electric charge moving in flat 
space-time of three dimensions is discussed. The divergences stemming from 
the pointness of the particle are studied. A consistent regularization 
procedure is proposed, which exploits the Poincar\'e invariance of the 
theory. Effective equation of motion of radiating charge in an external 
electromagnetic field is obtained via the consideration of energy-momentum 
and angular momentum conservation. This equation includes the effect of the
particle's own field. The radiation reaction is determined by the Lorentz 
force of point-like charge acting upon itself plus a non-local term which 
provides finiteness of the self-action.
\end{abstract}


\section{Introduction}
\setcounter{equation}{0}
There has been much recent interest \cite{Gl,KLS} in the renormalization 
procedure in classical electrodynamics of a point particle moving in flat 
space-time of arbitrary dimensions. The main task is to derive the analogue 
of the well-known Lorentz-Dirac equation \cite{Dir}. The Lorentz-Dirac 
equation is an equation of motion for a charged particle under the influence 
of an external force as well as its own electromagnetic field. (For a modern 
review see \cite{Rohr,TVW,PsPr}.)

A special attention in \cite{Gl,KLS} is devoted to the mass renormalization 
in $2+1$ theory. (Note that electrodynamics in Minkowski space 
${\mathbb M}_{\,3}$ is quite different from the conventional $3+1$ 
electrodynamics where one space dimension is reduced because of symmetry of 
the specific problem. For example, small charged balls on a plane are 
interacted inversely with the square of the distance between them, while in 
${\mathbb M}_{\,3}$ the Coulomb field of a small static charged disk scales 
as $|{\bf r}|^{-1}$.) An essential feature of $2+1$ electrodynamics is that 
the Huygens principle does not hold and radiation develops a tail, as it is 
in curved space-time of four dimensions \cite{WB} where electromagnetic 
waves propagate not just at the speed of light, but all speeds smaller than 
or equal to it.

In \cite{Gl,KLS} the self-force on a point-like particle is calculated 
from the local fields in the immediate vicinity of its trajectory. The 
schemes involve some prescriptions for subtracting away the infinite 
contributions to the force due to the singular nature of the field on the 
particle's world line. In \cite{KLS} the procedure of regularization is 
based on the methods of functional analysis which are applied to the Taylor 
expansion of the retarded Green's function. The authors derive the 
covariant analogue of the Lorentz-Dirac equation which is something other 
than that obtained in \cite{Gl}. Both the divergent self-energy absorbed by 
``bare'' mass of point-like charge, and the radiative term which leads an 
independent existence, are non-local. (They depend not only on the current 
state of motion of the particle, but also on its past history.)

In this paper we develop a consistent regularization procedure which 
exploits the symmetry properties of $2+1$ electrodynamics. It can be 
summarized as a simple rule which obeys the spirit of Dirac scheme of 
decomposition of vector potential of a point-like charge.

According to the scheme proposed by Dirac in his seminal paper 
\cite{Dir}, one can decompose retarded Green's function associated with 
the four-dimensional Maxwell field equation $G^{ret}(x,z)=G^{sym}(x,z)+
G^{rad}(x,z)$. The first term, $G^{sym}(x,z)$, is one-half sum of the 
retarded and the advanced Green's functions; it is just singular as 
$G^{ret}(x,z)$. The second one, $G^{rad}(x,z)$, is one-half of the retarded 
minus one-half of the advanced Green's functions; it satisfies the 
homogeneous wave equation. Convolving the source with the Green's functions
$G^{sym}(x,z)$ and $G^{rad}(x,z)$ yield the singular and the radiative 
parts of vector potential of a point-like charge, respectively.

The analogous decomposition of Green's function in curved spacetime is 
much more delicate because of richer causal structure. Detweiler and 
Whiting \cite{DW} modify the singular Green's function by means of a
two-point function $v(x,z)$ which is symmetric in its arguments. It is 
constructed from the solutions of the homogeneous wave equation in such a 
way that a new symmetric Green's function 
$G^S(x,z)=G^{sym}(x,z)+1/(8\pi)v(x,z)$ has no support within the null cone.

The physically relevant solution to the wave equation is obviously the 
retarded solution. In \cite{Teit} the Lorentz-Dirac equation is derived 
within the framework of retarded causality. Teitelboim substitutes the 
retarded Li\'enard-Wiechert fields in the electromagnetic field's 
stress-energy tensor. The author computes the flow of energy-momentum which 
flows across a tilted hyperplane which is orthogonal to particle's 
four-velocity at the instant of observation. The effective equation of motion is 
obtained in \cite{Teit} via consideration of energy-momentum conservation. 
Similarly, L\'opez and Villarroel \cite{LV} find out the total angular 
momentum carried by electromagnetic field of a point-like charge. 

Outgoing waves carry energy-momentum and angular momentum; the radiation 
removes energy, momentum, and angular momentum from the source which then 
undergoes a radiation reaction. It is shown \cite{Yar03} that the 
Lorentz-Dirac equation can be derived from the energy-momentum and angular 
momentum balance equations. In ref.\cite{Yar04} the analogue of the 
Lorentz-Dirac equation in six dimensions is obtained via analysis of 21 
conserved quantities which correspond to the symmetry of an isolated point 
particle coupled with electromagnetic field. (First, this equation was 
obtained by Kosyakov in \cite{Kos} via the consideration of energy-momentum 
conservation. An alternative derivation was produced by Kazinski, Lyakhovich 
and Sharapov in \cite{KLS}.)

In non-local theories, the computation of Noether quantities is highly 
nontrivial. Quinn and Wald \cite{QW} study the energy-momentum conservation 
for point charge moving in curved spacetime. The Stokes' theorem is applied 
to the integral of flux of electromagnetic energy over the compact region 
$V(t^+,t^-)$. It is expanded to the limits $t^-\to -\infty$ and 
$t^+\to +\infty$, so that finally the boundary of the integration 
domain involves smooth spacelike hypersurfaces at the remote past and in 
the distant future. The spacetime is asymptotically flat here. The authors 
prove that the net energy radiated to infinity is equal to the total work 
done on the particle by the electromagnetic self-force. (DeWitt-Brehme 
\cite{WB,Hb} radiation-reaction force is meant.) It is shown also \cite{QW} 
that the total work done by the gravitational self-force is equal to the 
energy radiated (in gravitational waves) by the particle. (The effective 
equation of motion of a point mass undergoing radiation reaction is 
obtained in \cite{MST}; see also review \cite{Pois} where the motion of a 
point electric charge, a point scalar charge and a point mass in curved 
spacetime is considered in detail.)

In the present paper, we calculate the total flows of energy-momentum and 
angular momentum of the retarded field which flow across a plane 
$\Sigma_t=\{y\in{\mathbb M}_{\,3}:y^0=t\}$ associated with an unmoving 
observer. It is organized as follows. In Section \ref{field}, we 
recall the retarded and the advanced Green's functions associated with the 
three-dimensional D'Alembert operator. Convolving them with the point 
source, we derive the retarded and the advanced vector potential and field 
strengths. In the \ref{trace}, we trace a series of stages in calculation 
of surface integral which gives the energy-momentum carried by the retarded 
electromagnetic field. We integrate the Maxwell energy-momentum tensor 
density over the variables which parametrize the surface of integration 
$\Sigma_t$. Resulting expression becomes a combination of two-point 
functions depending on the state of particle's motion at instants $t_1$ and 
$t_2$ before the observation instant $t$. They are integrated over 
particle's world line twice. We arrange them in Section \ref{Noe}. We split 
the momentum three-vector carried by electromagnetic field into singular and 
radiative parts by means of Dirac scheme which deals with the fields taken 
on the world line only. All diverging quantities have disappeared into the 
procedure of mass renormalization while radiative terms lead independent 
existence. In analogous way we analyze the angular momentum of 
electromagnetic field. Total energy-momentum and total angular momentum of 
our particle plus field system depend on already renormalized particle's 
individual characteristics and {\it radiative} parts of Noether quantities. 
In Section \ref{meq}, we derive the effective equation of motion of 
radiating charge via analysis of balance equations. In Section \ref{concl}, 
we discuss the result and its implications.

\section{Electromagnetic field in 2+1 theory}\label{field}
\setcounter{equation}{0}
We consider an electromagnetic field produced by a particle with 
$\delta$-shaped distribution of the electric charge $e$ moving on a world 
line $\zeta\subset {\mathbb M}_{\,3}$ described by functions $z^\mu(\tau)$ 
of proper time $\tau$. The Maxwell equations
\begin{equation}\label{Me}
F^{\alpha\beta}{}_{,\beta}=2\pi j^\alpha,
\end{equation}
where current density $j^\alpha$ is given by
\begin{equation}\label{je}
j^\alpha=e\int_{-\infty}^{+\infty}{\rm d}\tau 
u^\alpha (\tau)\delta^{(3)}(y-z(\tau)),
\end{equation} 
governs the propagation of the electromagnetic field. $u^\alpha (\tau)$ 
denotes the (normalized) three-velocity vector $dz^\alpha(\tau)/d\tau$ and 
$\delta^{(3)}(y-z)=\delta(y^0-z^0)\delta(y^1-z^1)\delta(y^2-z^2)$ is a 
three-dimensional Dirac distribution supported on the particle's world 
line $\zeta$. Both the strength tensor $F^{\alpha\beta}$ and the 
current density $j^\alpha$ are evaluated at a field point $y\in{\mathbb 
M}_{\,3}$.

We express the electromagnetic field in terms of a vector potential, 
$\hat F=d\hat A$. We impose the Lorentz gauge $A^\alpha{}_{,\alpha}=0$; then 
the Maxwell field equation (\ref{Me}) becomes
\begin{equation} 
\square A^\alpha =-2\pi j^\alpha .
\end{equation} 
In $2+1$ theory, the retarded Green's function associated with the 
D'Alembert operator 
$\square:=\eta^{\alpha\beta}\partial_\alpha\partial_\beta$ has support not 
just on the future light cone of the emission point $x$, but extends inside 
the light cone as well \cite{Gl,KLS}:
\begin{equation} \label{G3}
G_{2+1}^{ret}(y,z)=\frac{\theta(y^0-x^0-|{\mathbf y}-{\mathbf 
x}|)}{\sqrt{-(y-x)^2}}.
\end{equation}
$\theta(y^0-x^0-|{\mathbf y}-{\mathbf x}|)$ is a step function defined to be 
one if $y^0-x^0\ge |{\mathbf y}-{\mathbf x}|$, and defined to be zero 
otherwise.

Convolving the retarded Green function (\ref{G3}) with charge-current 
density $j^\alpha(x)$, we construct the retarded Li\'enard-Wiechert 
potential $A_\mu^{ret}(y)$ in three dimensions. It is generated by the point 
charge during its entire past history before the retarded time 
$\tau^{ret}(y)$ associated with the field point $y$. Apart from non-local 
term 
\begin{equation}\label{Fth}
F_{\mu\nu}^{(\theta)}=-e\int_{-\infty}^{\tau^{ret}(y)}{\rm d}\tau 
\frac{u_\mu K_\nu - u_\nu K_\mu}{[-(K\cdot K)]^{3/2}},
\end{equation}
the strength tensor $F_{\mu\nu}^{ret}=\partial_\mu A^{ret}_\nu 
-\partial_\nu A^{ret}_\mu$ of the adjunct electromagnetic field
contains also local term 
\begin{equation}\label{Fd}
F_{\mu\nu}^{(\delta)}=\lim_{\tau\to\tau^{ret}}\frac{e}{\sqrt{-(K\cdot 
K)}}\frac{u_\mu K_\nu - u_\nu K_\mu}{-(K\cdot u)}.
\end{equation}
which is due to differentiation of $\theta$-function involved in 
$A_\mu^{ret}(y)$. (By $K^\mu=y^\mu - z^\mu(\tau)$ we denote the unique 
timelike (or null) vector pointing from the emission point 
$z(\tau)\in\zeta$ to a field point $y$.) 

The terms separately diverge on the light cone. But the singularity, 
however, can be removed from the sum of ${\hat F}^{(\delta)}$ and ${\hat 
F}^{(\theta)}$. Using the identity
\begin{equation}\label{Ku}
\frac{1}{[-(K\cdot K)]^{3/2}}=\frac{1}{-(K\cdot 
u)}\frac{d}{d\tau}\frac{1}{\sqrt{-(K\cdot K)}}
\end{equation}
in equation (\ref{Fth}) yields
\begin{eqnarray}\label{Ft}
F_{\mu\nu}^{(\theta)}&=&-\frac{e}{\sqrt{-(K\cdot K)}}\left.
\frac{u_\mu K_\nu - u_\nu K_\mu}{-(K\cdot u)}
\right|_{\tau\to -\infty}^{\tau\to\tau^{ret}(y)}
\\
&+&e\int_{-\infty}^{\tau^{ret}(y)}
\frac{{\rm d}\tau}{\sqrt{-(K\cdot K)}}\left\{
\frac{a_\mu K_\nu - a_\nu K_\mu}{-(K\cdot u)}
+\frac{u_\mu K_\nu - u_\nu K_\mu}{[-(K\cdot u)]^2}\left[1+(K\cdot a)\right]
\right\}\nonumber
\end{eqnarray}
after integration by parts. Taking into account that $1/\sqrt{-(K\cdot K)}$ 
vanishes whenever $\tau\to -\infty$\footnote{We assume that average 
velocities are not large enough to initiate particle creation and 
annihilation, so that ``space contribution'' $|{\mathbf K}|$ cannot match 
with an extremely large zeroth component $K^0$.}, we finally obtain the 
expression
\begin{equation}\label{Fret}
{\hat F}^{ret}(y)= e\int_{-\infty}^{\tau^{ret}(y)}
\frac{{\rm d}\tau }{\sqrt{-(K\cdot K)}}\left\{
\frac{a\wedge K}{r}
+\frac{u\wedge K}{r^2}\left[1+(K\cdot a)\right]
\right\}
\end{equation}
which is regular on the light cone. It diverges on the particle's 
trajectory only. Symbol $\wedge$ denotes the wedge product. In $2+1$ 
electrodynamics, outgoing waves propagate not just at the speed of light, 
but all speeds smaller than or equal to it. The invariant quantity
\begin{equation}\label{rint}
r=-(K\cdot u)
\end{equation}
is an affine parameter on the time-like (null) geodesic that links $y$ to 
$z(\tau)$; it can be loosely interpreted as the time delay between $y$ and
$z(\tau)$ as measured by an observer moving with the particle. Because the 
speed of light is set to unity, parameter $r(\tau^{ret})$ is also the 
spatial distance between $z(\tau^{ret})$ and $y$ as measured in this 
momentarily comoving Lorentz frame.

In three dimensions, the advanced Green's function is nonzero in the past of 
the emission point $x$. The advanced strength tensor
\begin{equation}\label{Fadv}
{\hat F}^{adv}(y)= e\int^{+\infty}_{\tau^{adv}(y)}
\frac{{\rm d}\tau }{\sqrt{-(K\cdot K)}}\left\{
\frac{a\wedge K}{r}
+\frac{u\wedge K}{r^2}\left[1+(K\cdot a)\right]
\right\}
\end{equation}
is generated by the point charge during its entire future history following 
the advanced time associated with $y$.

\section{Bound and radiative parts of Noether quantities}\label{Noe}
\setcounter{equation}{0}

Equation of motion of radiating charge will be derived in the following
section via analysis of the total flows of (retarded) electromagnetic field 
energy-momentum and angular momentum across a hyperplane 
$\Sigma_t=\{y\in{\mathbb M}_{\,3}:y^0=t\}$. Noether quantities are given by 
surface integrals:
\begin{eqnarray}\label{pem}
p^\nu_{\rm em}(t)&=&\int_{\Sigma_t} {\rm d}\sigma_0 T^{0\nu}\\
M^{\mu\nu}_{\rm em}(t)&=&\int_{\Sigma_t} 
{\rm d}\sigma_0\left(y^\mu T^{0\nu} - y^\nu T^{0\mu}\right).
\end{eqnarray}
The electromagnetic field's stress-energy tensor $\hat T$ has the components
\begin{equation}\label{T}
2\pi T^{\mu\nu}=F^{\mu\lambda}F^\nu{}_{\lambda}- 
1/4\eta^{\mu\nu}F^{\kappa\lambda}F_{\kappa\lambda}
\end{equation}
where $\hat F$ is the non-local strength tensor (\ref{Fret}).

The computation is not a trivial matter, since the Maxwell energy-momentum 
tensor density evaluated at field point $y\in\Sigma_t$ is non-local. In odd 
dimensions, the retarded field is generated by the portion of the world line 
$\zeta$ that corresponds to the particle's history {\it before} 
$t^{ret}(y)$. Since the stress-energy tensor is quadratic in field 
strengths, we should {\it twice} integrate it over $\zeta$. We integrate it 
also over two variables which parametrize $\Sigma_t$ in order to calculate 
energy-momentum and angular momentum which flow across this plane. The 
integrand describes the combination of outgoing electromagnetic waves 
emitted at instants $t_1$ and $t_2$ {\it before} $t$. (The situation is 
pictured in figure \ref{f_ab}.) We trace a series of stages in calculations 
in the Appendix; a detailed description is published \cite{YJMP}. To 
summarize briefly, the integration of the energy-momentum and angular 
momentum tensor densities over variables which parametrize $\Sigma_t$ 
results twofold integrals, we obtain two-point and three-point functions 
defined on the world line only, which are twice integrated over $\zeta$.
(See equations (\ref{p0em})-(\ref{sing}) in the \ref{trace} where the 
integration of the energy-momentum tensor density in 2+1 dimensions is 
considered.)

\begin{figure}
\begin{center}
\epsfclipon
\epsfig{file=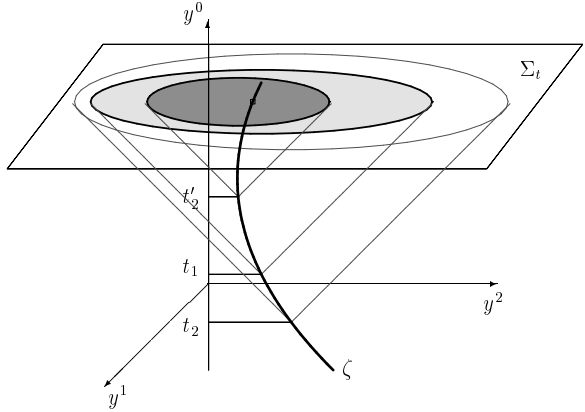,width=8cm}
\end{center}
\caption{\label{f_ab}
\small The integration of energy and momentum densities over 
two-dimensional plane $y^0=t$ means the study of interference of 
outgoing electromagnetic waves generated by different points on particle's 
world line. Let the instant $t_1$ be fixed. Outgoing electromagnetic waves 
generated by the portion of the world line that corresponds to the interval 
$-\infty <t_2<t_1$ combine within the gray disc with radius $k_1^0=t-t_1$. 
If the domain of integration $t_1<t_2\le t$, the waves joint together inside 
the dark disc with radius $k_2^0=t-t_2'$. 
}
\end{figure}

All the three-point terms which contain the observation instant $t$ belong 
to the (diverging) bound parts of Noether quantities. They are permanently 
``attached'' to the charge and are carried along with it. The radiation 
parts contain the two-point functions depending on particle's position and 
velocity referred to the instants $t_1$ and $t_2$ before $t$.

Having done reparametrization, we rewrite the two-point functions which 
arise in energy-momentum and angular momentum in a manifestly covariant 
fashion:
\begin{eqnarray}\label{pm12}
p^\mu_{12}&=&u_{1,\alpha}\frac{-u_2^\alpha q^\mu +u_2^\mu 
q^\alpha}{[-(q\cdot q)]^{3/2}}\nonumber\\
m^{\mu\nu}_{12}&=&z^\mu_1p^\nu_{12} - z^\nu_1p^\mu_{12}.
\end{eqnarray}
Index $1$ indicates that the particle's velocity or position is referred to the 
instant $\tau_1\in ]-\infty,\tau]$ while index $2$ says that the particle's 
characteristics are evaluated at instant $\tau_2\le\tau_1$. Here, 
$q^\mu=z_1^\mu-z_2^\mu$ defines the unique timelike geodesic connecting a 
field point $z(\tau_1)\in\zeta$ to an emission point
$z(\tau_2)\in\zeta$. Since $q^\mu(\tau_2,\tau_1)=-q^\mu(\tau_1,\tau_2)$, the 
reciprocity relations are satisfied:
\begin{equation}\label{rp} 
\left.p^\mu_{21}\right|_{1\leftrightarrow 2}=-p^\mu_{12},\qquad 
\left.m^{\mu\nu}_{21}\right|_{1\leftrightarrow 2}=-m^{\mu\nu}_{12}.
\end{equation}

The radiative parts of electromagnetic field's energy-momentum and 
angular momentum are as follows:
\begin{eqnarray}\label{prom}
p^\mu_{\rm R}(\tau)&=&\frac{e^2}{2}
\int_{-\infty}^\tau {\rm d}\tau_1\int_{-\infty}^{\tau_1}{\rm d}\tau_2
\left(p^\mu_{12}+p^\mu_{21}\right)\\\label{M_R}
M^{\mu\nu}_{\rm R}(\tau)&=&\frac{e^2}{2}
\int_{-\infty}^\tau {\rm d}\tau_1\int_{-\infty}^{\tau_1}{\rm d}\tau_2
\left(m^{\mu\nu}_{12}+m^{\mu\nu}_{21}\right).
\end{eqnarray}
Our next task will be to elucidate geometrical sense of these expressions.

\begin{figure}
\begin{center}
\epsfclipon
\epsfig{file=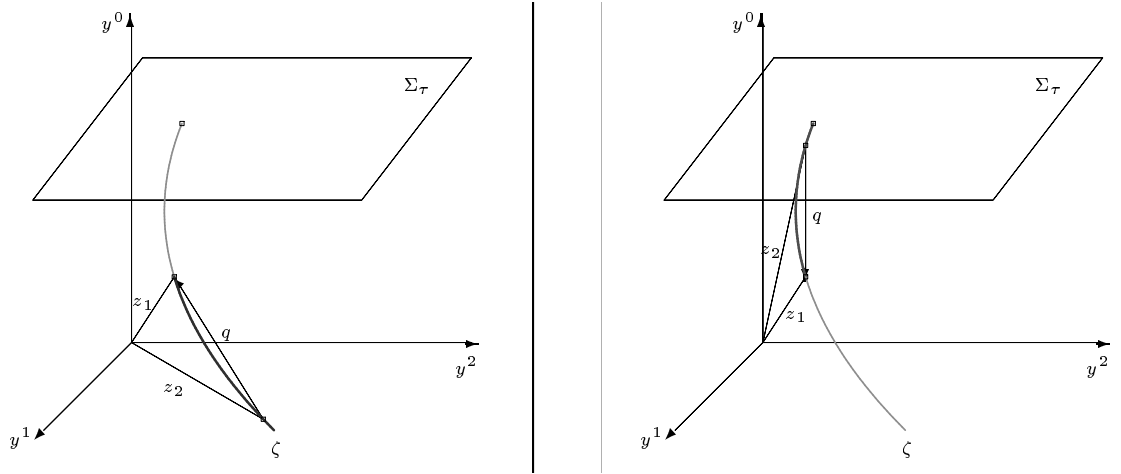,width=15cm}
\end{center}
\caption{\label{Dscheme}
\small The term (\ref{Gret}) with integration over the portion of the 
world line {\it before} $\tau_1$ we call ``retarded''. The term (\ref{Gadv}) 
with integration over the portion of the world line {\it after} $\tau_1$ 
we call ``advanced''. For an observer placed at point $z(\tau_1)\in\zeta$, 
the regular part (\ref{pR}) of Noether quantities carried by electromagnetic 
field looks as the combination of incoming and outgoing radiation. 
And yet the retarded causality is not violated. We still consider the 
interference of outgoing waves presented at the observation plane 
$\Sigma_\tau$. The electromagnetic field carries information about the 
charge's past.
}
\end{figure}

We take the first terms under the integral signs in equations  (\ref{prom}),
(\ref{M_R}) and denote them as $g^\alpha_{12}=(p^\mu_{12},m^{\mu\nu}_{12})$. 
We integrate the two-point functions (\ref{pm12}) over the portion of the 
world line which corresponds to the interval $-\infty<\tau_2\le\tau_1$. We 
introduce the functions:
\begin{equation}\label{Gret}
G_{\rm ret}^\alpha 
(\tau_1)=e^2\int_{-\infty}^{\tau_1}{\rm d}\tau_2 g^\alpha_{12}.
\end{equation}
Next we take the remaining terms $g^\alpha_{21}$; we change the order of 
integration over
the domain $D_\tau=\{(\tau_1,\tau_2)\in{\mathbb R}^{\,2}:
\tau_1\in ]-\infty,\tau],\tau_2\le\tau_1\}$:
\begin{equation}
\int_{-\infty}^\tau {\rm 
d}\tau_1\int_{-\infty}^{\tau_1}{\rm d}\tau_2g^\alpha_{21}
=\int_{-\infty}^\tau {\rm d}\tau_2\int_{\tau_2}^\tau 
{\rm d}\tau_1g^\alpha_{21}.
\end{equation}
Since instants $\tau_1$ and $\tau_2$ label points at the same world line 
$\zeta$, one can interchange indices ``first'' and ``second'' on the 
right-hand side of this equation. Taking into account relations (\ref{rp}), 
we finally obtain:
\begin{equation}
\int_{-\infty}^\tau {\rm 
d}\tau_1\int_{-\infty}^{\tau_1}{\rm d}\tau_2g^\alpha_{21}=
-\int_{-\infty}^\tau {\rm d}\tau_1\int_{\tau_1}^\tau 
{\rm d}\tau_2g^\alpha_{12}.
\end{equation}
The integrand coincides with that under integral sign on the right-hand side 
of equation (\ref{Gret}) while the domain of inner integration corresponds 
to the interval $\tau_1\le\tau_2\le\tau$. We introduce the function
\begin{equation}\label{Gadv}
G_{\rm adv}^\alpha(\tau_1,\tau)= 
e^2\int_{\tau_1}^{\tau}{\rm d}\tau_2g^\alpha_{12}.
\end{equation}
We see that  the double integral in equation (\ref{prom}) can be expressed 
as one-half of $G_{\rm ret}$ minus one-half of $G_{\rm adv}$ integrated over 
the world line $\zeta$:
\begin{equation}\label{pR}
G^\alpha_{\rm R}(\tau)=\frac12\int_{-\infty}^\tau {\rm d}\tau_1\left[
G_{\rm ret}^\alpha(\tau_1) - G_{\rm adv}^\alpha(\tau,\tau_1)
\right].
\end{equation}
The situation is pictured in figure \ref{Dscheme}.

We evaluate the short-distance behaviour of the expression under the double 
integral in equation(\ref{pR}). Let $\tau_1$ be fixed and 
$\tau_1-\tau_2:=\Delta$ be a small parameter. With a degree of accuracy 
sufficient for our purposes
\begin{eqnarray}
\sqrt{-(q\cdot q)}&=&\Delta,\\
q^\mu&=&\Delta\left[u_1^\mu-a_1^\mu\frac{\Delta}{2}+
{\dot a}_1^\mu\frac{\Delta^2}{6}\right],\nonumber\\
u_2^\mu&=&u_1^\mu-a_1^\mu\Delta+{\dot a}_1^\mu\frac{\Delta^2}{2}.\nonumber
\end{eqnarray}
Substituting these into integrand of the double integral of equation 
(\ref{pR}) and passing to the limit $\Delta\to 0$ yields regular expression
\begin{equation}\label{lim}
\lim_{\tau_2\to\tau_1}\left[
\frac12u_{1,\alpha}\frac{-u_2^\alpha q^\mu +u_2^\mu 
q^\alpha}{[-(q\cdot q)]^{3/2}}+
\frac12u_{2,\alpha}\frac{-u_1^\alpha q^\mu +u_1^\mu 
q^\alpha}{[-(q\cdot q)]^{3/2}}
\right]
=\frac13(a_1)^2u_1^\mu-\frac{1}{12}{\dot a}_1^\mu.
\end{equation}
Therefore the subscript ``R'' stands for ``regular'' as well as 
``radiative''.

Alternatively, choosing the linear superposition
\begin{equation}\label{pS}
G^\alpha_{\rm S}(\tau)=
\frac12\int_{-\infty}^\tau {\rm d}\tau_1\left[
G_{\rm ret}^\alpha(\tau_1) + G_{\rm adv}^\alpha(\tau,\tau_1)
\right]
\end{equation}
we restore the singular parts of Noether quantities:
\begin{eqnarray}\label{pMS}
p^\mu_{\rm S}(\tau)&=&\frac{e^2}{2}
\int_{-\infty}^\tau 
{\rm d}\tau_2\frac{u^\mu(\tau_2)}{\sqrt{-(q\cdot q)}},
\\
M^{\mu\nu}_{\rm S}(\tau)&=&\frac{e^2}{2}
\int_{-\infty}^\tau {\rm d}\tau_2
\frac{z^\mu(\tau)u^\nu(\tau_2)
-z^\nu(\tau)u^\mu(\tau_2)}{\sqrt{-(q\cdot q)}}.
\nonumber
\end{eqnarray}
Since this non-local term diverges, the subscript ``S'' stands for 
``singular'' as well as ``symmetric''. 

\section{Equation of motion of radiating charge}\label{meq}
\setcounter{equation}{0}

We therefore introduce the radiative part $p_{\rm R}$ of energy-momentum 
(see equation (\ref{prom})) and postulate that it, and it alone, exerts a 
force on the particle. Singular part, $p_{\rm S}$, should be coupled with 
particle's three-momentum, so that ``dressed'' charged particle would not 
undergo any additional radiation reaction. Already renormalized particle's 
individual three-momentum, say $p_{\rm part}$, together with $p_{\rm R}$ 
constitute the total energy-momentum of our composite particle plus field 
system:  $P=p_{\rm part}+p_{\rm R}$. Since $P$ does not change with time, 
its time derivative yields 
\begin{eqnarray}\label{pdot}
{\dot p}^\mu_{\rm part}(\tau)&=&-{\dot p}^\mu_{\rm R}\\
&=&-\frac{e^2}{2}\int_{-\infty}^\tau {\rm d} s
\left[
u_{\tau,\alpha}\frac{-u_s^\alpha q^\mu +u_s^\mu
q^\alpha}{[-(q\cdot q)]^{3/2}}+
u_{s,\alpha}\frac{-u_\tau^\alpha q^\mu +u_\tau^\mu 
q^\alpha}{[-(q\cdot q)]^{3/2}}
\right].\nonumber
\end{eqnarray} 
(The overdot means the derivation with respect to proper time $\tau$.)
Here index $\tau$ indicates that particle's velocity or position is 
referred to the observation instant $\tau$ while index 
$s$ says that the particle's characteristics are evaluated at instant 
$s\le\tau$.

Our next task is to derive expression which explain how three-momentum of 
``dressed'' charged particle depends on its individual characteristics 
(velocity, position, mass etc). We do not make any assumptions about the 
particle structure, its charge distribution and its size. We only assume 
that the particle 3-momentum $p_{\rm part}$ is finite. To find out 
the desired expression, we analyze conserved quantities corresponding to the 
invariance of the theory under proper homogeneous Lorentz transformations. 
The total angular momentum, say $M$, consists of particle's 
angular momentum $z\wedge p_{\rm part}$ and radiative part (\ref{M_R}) of 
angular momentum carried by electromagnetic field:
\begin{equation}\label{Mtot}
M^{\mu\nu}=z_\tau^\mu p_{\rm part}^\nu(\tau) 
- z_\tau^\nu p_{\rm part}^\mu(\tau) + M^{\mu\nu}_{\rm R}(\tau).
\end{equation}

Having differentiated (\ref{Mtot}) and inserting equation (\ref{pdot}),
we arrive at the equality
\begin{equation}\label{Mdot}
u_\tau\wedge p_{\rm part}=
\frac{e^2}{2}\int_{-\infty}^\tau {\rm d} s\frac{u_\tau\wedge 
u_s}{\sqrt{-(q\cdot q)}}.
\end{equation}
Apart from usual velocity term, the simplest solution contains also 
singular contribution (\ref{pMS}) from particle's electromagnetic field:
\begin{eqnarray}\label{part0}
p_{\rm part}^\mu(\tau)&=&m_0u^\mu(\tau)+\frac{e^2}{2}\int_{-\infty}^\tau 
{\rm d} s\frac{u^\mu(s)}{\sqrt{-(q\cdot q)}}\nonumber\\
&=&m_0u^\mu(\tau)+p_S^\mu(\tau).
\end{eqnarray}
Near the coincidence limit $s\to\tau$ the denominator of the integrand 
in equation (\ref{part0}) behaves as $\tau -s$, so that integral diverges 
logarithmically. Since $p_{\rm part}$ is proclaimed to be finite, the 
``bare'' mass $m_0$ should absorb divergency within the renormalization 
procedure. To cancel the infinity Kazinski, Lyakhovich and Sharapov 
\cite{KLS} add the term which, in our notations, looks as follows:
\begin{equation}\label{dm}
\delta m=\frac{e^2}{2}\int_{-\infty}^\tau 
\frac{{\rm d} s}{\sqrt{-(q\cdot q)}}
\end{equation}
(see \cite[equation (38)]{KLS}). Taking it into account we obtain
\begin{equation}\label{part}
p_{\rm part}^\mu(\tau)=mu^\mu(\tau)+\frac{e^2}{2}\int_{-\infty}^\tau 
{\rm d} s\frac{u^\mu(s)-u^\mu(\tau)}{\sqrt{-(q\cdot q)}}
\end{equation}
where already renormalized mass $m$ is proclaimed to be finite.

The scalar product of particle three-velocity on the first-order 
time-derivative of particle three-momentum (\ref{pdot}) is as follows:
\begin{equation}\label{udp}
({\dot p}_{\rm part}\cdot u_\tau)=\frac{e^2}{2}\int_{-\infty}^\tau ds
\left[
(u_\tau\cdot u_s)\frac{(u_\tau\cdot q)}{[-(q\cdot q)]^{3/2}} +
\frac{(u_s\cdot q)}{[-(q\cdot q)]^{3/2}}
\right].
\end{equation}
Since $(u\cdot a)=0$, the scalar product of particle acceleration on the 
particle three-momentum (\ref{part}) is given by
\begin{equation}\label{ap}
(p_{\rm part}\cdot a_\tau)=\frac{e^2}{2}\int_{-\infty}^\tau ds
\frac{(a_\tau\cdot u_s)}{\sqrt{-(q\cdot q)}}.
\end{equation}
Summing up (\ref{udp}) and (\ref{ap}) we obtain
\begin{equation}\label{dpdp}
\frac{d}{d\tau}(p_{\rm part}\cdot u_\tau)=
\frac{e^2}{2}\int_{-\infty}^\tau ds\left\{
\frac{\partial}{\partial\tau}\left[\frac{(u_\tau\cdot 
u_s)}{\sqrt{-(q\cdot q)}}\right] +
\frac{(u_s\cdot q)}{[-(q\cdot q)]^{3/2}}
\right\}.
\end{equation}

Alternatively, the scalar product of 3-momentum (\ref{part}) and 3-velocity 
is as follows:
\begin{equation}\label{pu}
(p_{\rm part}\cdot u_\tau)=-m+\frac{e^2}{2}\int_{-\infty}^\tau ds
\frac{(u_\tau\cdot u_s)+1}{\sqrt{-(q\cdot q)}}.
\end{equation}
Further we compare its differential consequence with equation (\ref{dpdp}). 
A surprising feature of the already renormalized {\it dynamical} mass $m$ 
is that it depends on $\tau$: 
\begin{equation}\label{dotm}
\dot m=\frac{e^2}{2}\int_{-\infty}^\tau ds
\frac{(q\cdot u_\tau)-(q\cdot u_s)}{[-(q\cdot q)]^{3/2}}.
\end{equation}
It is interesting that similar phenomenon occurs in the theory which 
describes a point-like charge coupled with massless scalar field in flat 
spacetime of three dimensions \cite{Br}. The charge loses its mass through 
the emission of monopole radiation.

Having integrated derivative (\ref{dotm}) over the world line $\zeta$  we 
obtain
\begin{eqnarray}
m&=&m_0+\frac{e^2}{2}\int_{-\infty}^\tau 
{\rm d}\tau_1\int_{-\infty}^{\tau_1} {\rm d}\tau_2 
\left[
\frac{\partial}{\partial\tau_1}\left(\frac{1}{\sqrt{-(q\cdot q)}}\right)+
\frac{\partial}{\partial\tau_2}\left(\frac{1}{\sqrt{-(q\cdot q)}}\right)
\right]\nonumber\\
&=&m_0+\frac{e^2}{2}\int_{-\infty}^\tau \frac{{\rm d} s}{\sqrt{-(q\cdot q)}}
\end{eqnarray}
where $m_0$ is an infinite ``bare'' mass of the particle. Inserting this 
into (\ref{part}) we arrive at the equality $p_{\rm 
part}^\mu(\tau)=m_0u^\mu_\tau+p_S^\mu$ which shows that particle's 
momentum renormalization agrees with the renormalization of mass.

The main goal of the present paper is to compute the effective equation of 
motion of radiating charge in $2+1$ dimensions. To do it we replace ${\dot 
p}_{\rm part}^\mu$ on the left-hand side of equation (\ref{pdot}) by 
differential consequence of equation (\ref{part}) where the right-hand side 
of equation (\ref{dotm}) substitutes for $\dot m$. At the end of a 
straightforward calculations, we obtain 
\begin{equation}\label{me}
ma^\mu_\tau=\frac{e^2}{2}a^\mu_\tau
- e^2u_{\tau,\alpha}\int_{-\infty}^\tau {\rm d} s\frac{-u_s^\alpha q^\mu
+u_s^\mu q^\alpha}{[-(q\cdot q)]^{3/2}}
+\frac{e^2}{2}a^\mu_\tau\int_{-\infty}^\tau 
\frac{{\rm d} s}{\sqrt{-(q\cdot q)}}.
\end{equation}
The first term on the right-hand side of this equation looks horribly 
irrelevant. It arises also in \cite[equation (40)]{KLS} where it is called 
``the local part of Lorentz-Dirac (LD) force''. It is worth noting that 
Kazinski, Lyakhovich and Sharapov regularize the local part (\ref{Fd}) of 
the retarded electromagnetic field and its non-local part (\ref{Fth})  
separately. (More exactly, the authors manipulate with convolution of these 
terms with particle's velocity taken at instant $\tau$.) Since the local 
part is proportional to particle's acceleration, they move it to the 
left-hand side of equation (\ref{me}) where one-half of the squared charge 
is proclaimed to be absorbed by mass. The remaining non-local part (the 
second term in equation (\ref{me})) together with the third term of this 
equation constitute ``the renormalized LD force'' which alone determines the 
radiation reaction.

But the finite first term in equation (\ref{me}) is the integral part of 
Lorentz self-force and, therefore, can not be cancelled within the 
regularization procedure. According to section \ref{field}, the sum of 
local (\ref{Fd}) and non-local (\ref{Ft}) parts of electromagnetic field 
results the non-local expression (\ref{Fret}) for the retarded 
electromagnetic field. The field strengths at point $z(\tau)\in\zeta$ 
generated by the portion of the world line {\it before} the observation 
instant $\tau$ looks as follows:
\begin{eqnarray}\label{retF}
F^{\mu\alpha}_{\rm ret}(\tau)&=&
e\int_{-\infty}^\tau\frac{{\rm d} s}{\sqrt{-(q\cdot q)}}
\left\{\frac{u_s^\mu q^\alpha -u_s^\alpha 
q^\mu}{r^2}\left[1+(a_s\cdot q)\right]
+\frac{a_s^\mu q^\alpha -a_s^\alpha q^\mu}{r}
\right\}\\
&=&\int_{-\infty}^\tau {\rm d} sf^{\mu\alpha}(\tau,s).\nonumber
\end{eqnarray}
Its convolution with particle's velocity is equal to the combination of the 
first term and of the second term in the right-hand side of equation 
(\ref{me}):
\begin{equation}\label{GFret}
eu_{\tau,\alpha}F_{\rm ret}^{\mu\alpha}(\tau)=\frac{e^2}{2}a_\tau^\mu - 
e^2u_{\tau,\alpha}\int_{-\infty}^\tau {\rm d} s\frac{-u_s^\alpha q^\mu
+u_s^\mu q^\alpha}{[-(q\cdot q)]^{3/2}}.
\end{equation}
(It may be checked via using the equality (\ref{Ku}) and integration by 
parts.) This relation prompts that the retarded Lorentz self-force should be 
substituted for this combination. If an external electromagnetic field $\hat 
F_{\rm ext}$ is applied, the equation of motion of radiating charge in $2+1$ 
theory becomes
\begin{equation}\label{mext}
ma^\mu_\tau=eu_{\tau,\alpha} F^{\mu\alpha}_{\rm ret}(\tau)+ 
\frac{e^2}{2}a^\mu_\tau\int_{-\infty}^\tau 
\frac{{\rm d} s}{\sqrt{-(q\cdot q)}}
+eu_{\tau,\alpha} F^{\mu\alpha}_{\rm ext}.
\end{equation}
The non-local term in equation (\ref{mext}) which is proportional to 
particle's acceleration $a(\tau)$ arises also in \cite{KLS}. It gives rise 
the renormalization of mass and provides proper short-distance behaviour of 
the perturbations due to the particle's own field. If $s\to\tau$ the 
integrand tends to three-dimensional analog of the Abraham radiation 
reaction vector:
\begin{equation}
\lim_{s\to\tau}\left[
eu_{\tau,\alpha}f^{\mu\alpha}(\tau,s)+
\frac{e^2}{2}\frac{a^\mu_\tau}{\sqrt{-(q\cdot q)}}
\right]=\frac23e^2\left({\dot a}^\mu-a^2u^\mu\right).
\end{equation}
(All quantities on the right-hand side refer to the instant of observation 
$\tau$.)

If one moves the second term to the left-hand side of equation (\ref{mext}), 
they restore unphysical motion equation which follows from variational 
principle: it involves an infinite ``bare'' mass and divergent Lorentz 
self-force.

\section{Conclusions}\label{concl}

In the present paper, we calculate the total flows of (retarded) 
electromagnetic field energy, momentum and angular momentum which flow 
across the plane $\Sigma_t=\{y\in{\mathbb M}_{\,3}:y^0=t\}$. We integrate 
the stress-energy tensor over two variables which parametrize $\Sigma_t$. 
Thanks to integration we reduce the support of the retarded and the advanced 
Green's functions to particle's trajectory.

The Dirac scheme which manipulates fields {\it on the world line only} is 
the key point of investigation. By fields we mean the convolution 
$eu_\nu(\tau_1) F^{\nu\mu}_{(\theta)}$ of three-velocity and non-local 
part (\ref{Fth}) of the retarded strength tensor evaluated at point 
$z(\tau_1)\in\zeta$; the torque of this ``Lorentz $\theta$-force'' arises 
in electromagnetic field's total angular momentum. (Singular 
$\delta$-term (\ref{Fd}) is defined on the light cone; it is meaningless 
since both the field point, $z(t_1)$, and the emission point, $z(t_2)$, 
lie on the time-like world line). The retarded and the ``advanced''
quantities arise naturally. The retarded Lorentz self-force as well as 
its torque contain integration over the portion of the world line which 
corresponds to the interval $-\infty<t_2\le t_1$.  Domain of integration
of their ``advanced'' counterparts corresponds to the interval 
$t_1\le t_2\le t$.

Noether quantity $G^\alpha_{\rm em}$ carried by electromagnetic field 
consists of terms of two quite different types: (i) singular, 
$G^\alpha_{\rm S}$, which is permanently ``attached'' to the source and 
carried along with it; (ii) radiative, $G^\alpha_{\rm R}$, which 
detaches itself from the charge and leads an independent existence.
The former is the half-sum of the retarded and advanced expressions,
integrated over $\zeta$, while the latter is the integral of one-half of 
the retarded quantity minus one-half of the advanced one. Within the 
regularization procedure, the bound terms $G_{\rm S}^\alpha$ are coupled 
with the energy-momentum and angular momentum of ``bare'' source, so that 
already renormalized characteristics $G^\alpha_{\rm part}$ of charged 
particle are proclaimed to be finite. Noether quantities which are 
properly conserved become
$$
G^\alpha=G^\alpha_{\rm part}+G^\alpha_{\rm R}.
$$

The energy-momentum balance equations define the change of particle's 
three-momentum under the influence of an external electromagnetic field 
where loss of energy due to radiation is taken into account. The angular 
momentum balance equations explain how this already renormalized 
three-momentum depend on particle's individual characteristics. They 
constitute the system of three linear equations in three components of 
particle's momentum. Its rank is equal to 2, so that arbitrary scalar 
function arises naturally. It can be interpreted as a {\it dynamical} mass 
of ``dressed'' charge which is proclaimed to be finite. A surprising feature 
is that this mass depends on the particle's history before the instant of 
observation when the charge is accelerated. Already renormalized 
particle's momentum contains, apart from usual velocity term, also 
non-local contribution from point-like particle's electromagnetic field.

Having substituted this expression in the energy-momentum balance equations, 
we derive three-dimensional analogue of the Lorentz-Dirac equation 
$$
ma^\mu_\tau=eu_{\tau,\alpha} F^{\mu\alpha}_{\rm ret}(\tau)+ 
\frac{e^2}{2}a^\mu_\tau\int_{-\infty}^\tau 
\frac{{\rm d} s}{\sqrt{-(q\cdot q)}}
+eu_{\tau,\alpha} F^{\mu\alpha}_{\rm ext}.
$$
The loss of energy due to radiation is determined by the work 
done by the Lorentz force of point-like charge acting upon itself. Non-local 
term which is proportional to particle's acceleration provides finiteness 
of the self-action. It is intimately connected with the renormalization 
of mass. The third term describes influence of an external field.

In \cite{Gl,KLS} the Lorentz self-force is replaced with the second 
term of right-hand side of equation (\ref{me}) which does not possess quite 
clear physical sense. Besides, in \cite{Gl} the non-local term which gives 
rise to the infinite mass renormalization and provides proper short-distance 
behaviour is as follows:
$$
\frac{e^2}{2}a^\mu_\tau\int_{-\infty}^\tau 
\frac{{\rm d} s}{|\tau-s|}.
$$
As noted in \cite{KLS}, there is a little sense in finiteness near the 
coincidence limit $s\to \tau$ since the expression for radiation reaction 
is not invariant with respect to reparametrization.

In this paper, we develop convenient technique which allows us to integrate 
non-local stress-energy tensor over the spacelike plane. The next step 
will be to implement this strategy to a point particle coupled to 
massive scalar field following an arbitrary trajectory on a flat 
spacetime. The Klein-Gordon field generated by the scalar charge holds 
energy near the particle. This circumstance makes unclear the procedure of 
decomposition of the energy-momentum into bound and radiative parts.

In \cite{AHNS} the remarkable correspondence is established between 
dynamical equations which govern behaviour of superfluid 
$\!\!\!\phantom{I}^4$He films and Maxwell equations for electrodynamics 
in $2+1$ dimensions (see also refs.\cite{FL,Z})\footnote{I wish to 
thank O.Derzhko for drawing these papers to my attention.}. Perhaps the 
effective equation of motion (\ref{mext}) will be useful in the study of 
phenomena in superfluid dynamics which correspond to the radiation 
friction in $2+1$ electrodynamics.

\section*{Acknowledgments}
I am grateful to Professor V.Tre\-tyak for continuous encouragement and for 
a helpful reading of this manuscript. I would like to thank A.Du\-vi\-ryak 
and R.Matsyuk for many useful discussions.


\section*{Appendix. Energy-momentum of electromagnetic field in 2+1 
dimensions}\label{trace}
\setcounter{equation}{0}

In this section, we trace a series of stages in calculation of the surface 
integral 
\begin{equation}\label{pem0}
p^\nu_{\rm em}(t)=\int_{\Sigma_t} {\rm d}\sigma_0 T^{0\nu}
\end{equation}
which gives the energy-momentum carried by electromagnetic field of a 
point-like source arbitrarily moving in ${\mathbb M}_{\,3}$.
Calculation of total angular momentum is virtually identical to that 
presented below, and we shall not bother with the details.

Huygens principle does not hold in three dimensions: a point 
$z(t_1)\in\zeta$ produces the disc of radius $t-t_1$ in the observation 
plane $\Sigma_t=\{y\in{\mathbb M}_3: y^0=t\}$ (see figure \ref{f_ab}). The 
integration of energy and momentum densities over $\Sigma_t$ means the study 
of interference of outgoing electromagnetic waves emitted by different 
points on $\zeta$:
\begin{equation}\label{pint}
p_{\rm 
em}^\alpha=\int\limits_{-\infty}^t{\rm d} t_1\int\limits_{-\infty}^{t_1}{\rm d} 
t_2
\int\limits_{0}^{k_1^0}{\rm d} R 
\int\limits_0^{2\pi}{\rm d}\varphi Jt^{0\alpha}_{12}
+
\int\limits_{-\infty}^t{\rm d} 
t_1\int\limits_{t_1}^t{\rm d} t_2\int\limits_{0}^{k_2^0}{\rm d} R
\int\limits_0^{2\pi}{\rm d}\varphi Jt^{0\alpha}_{12}\nonumber
\end{equation}
The first multiple integral calculates the interference of the disc 
emanated by fixed point $z(t_1)\in\zeta$ with radiation generated by 
all the points {\it before} the instant $t_1$. The second fourfold 
integral gives the contribution of points {\it after} $t_1$.

The integrand
\begin{equation}
2\pi t^{\alpha\beta}_{12}=
f_{(1)}^{\alpha\lambda}f_{(2)\lambda}^\beta -\frac14\eta^{\alpha\beta}
f_{(1)}^{\mu\nu}f^{(2)}_{\mu\nu}
\end{equation}
describes the combination of field strength densities at $y\in\Sigma_t$
\begin{equation}\label{F1F2}
{\hat f}_{(a)}(y)=\frac{e}{\sqrt{-(K_a\cdot K_a)}}\left(
\frac{{\dot v}_a\wedge K_a}{r_a}+\frac{v_a\wedge K_a}{(r_a)^2}
\left[\gamma_a^{-2}+(K_a\cdot {\dot v}_a)\right]
\right)
\end{equation}
generated by emission points $z(t_1)\in\zeta$ and $z(t_2)\in\zeta$.
The Jacobian $J$ corresponds to coordinate transformation
\begin{eqnarray}\label{ct}
y^0&=&t\\
y^i&=&\alpha z^i(t_1)+\beta z^i(t_2) + R\omega^i{}_j n^j\nonumber
\end{eqnarray} 
where $\alpha +\beta =1$ and $n^j=(\cos\varphi,\sin\varphi)$. Orthogonal 
matrix 
\begin{equation}\label{om}
\omega=\left(
\begin{array}{cc}
n_q^1& -n_q^2\\[0.5ex]
n_q^2& n_q^1
\end{array}
\right)
\end{equation}
rotates space axes till new $y^1-$axis be directed along two-vector ${\bf 
q}:={\bf z}(t_1)-{\bf z}(t_2)$. (We denote $n_q^i=q^i/q$.)

It is worth noting that time variables $t_1$ and $t_2$ parametrize the 
same world line $\zeta$. Coordinate transformation (\ref{ct}) is 
invariant with respect to the following reciprocity:
\begin{equation}
\Upsilon: t_1\leftrightarrow t_2, \alpha\leftrightarrow\beta, 
\varphi\mapsto\varphi+\pi.
\end{equation}
This symmetry provides identity of domains of fourfold integrals in 
energy-momentum (\ref{pint}).

\begin{figure}
\begin{center}
\epsfclipon
\epsfig{file=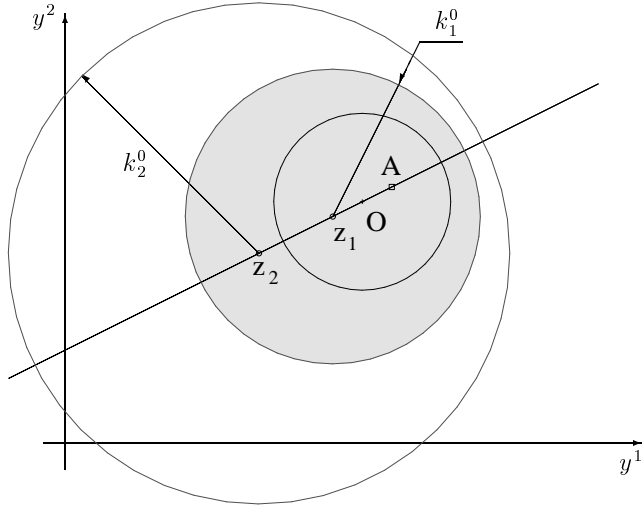,width=9cm}
\end{center}
\caption{\label{bt}
\small The interference picture in the plane $\Sigma_t$. The points 
$z(t_1)\in\zeta$ and $z(t_2)\in\zeta, t_2<t_1,$ emanate the radiation which 
filled up the discs centred at $z_1$ and $z_2$, respectively. The 
gray disc with radius $k_1^0=t-t_1$ is filled up by non-concentric 
circles centred at the line crossing both the points $z_1$ and $z_2$. 
If parameter $\beta$ vanishes the circle is centred at $z_1$, its 
radius is equal to $k_1^0$. If $\beta=\beta_0<0$, the circle reduces to the 
point $A$ labelled by the box symbol. In case of intermediate value 
$\beta_0<\beta<0$ we have the circle of radius $R$ with center 
at point $O$ between $z_1$ and $A$.}
\end{figure}

It is obvious that the support of double integral 
$\int_{-\infty}^t{\rm d} t_1\int_{t_1}^{t}{\rm d} t_2$ coincides with the support 
of the integral $\int_{-\infty}^t{\rm d} t_2\int_{-\infty}^{t_2}{\rm d} t_1$. 
Since instants $t_1$ and $t_2$ label different points at the same world line 
$\zeta$, one can interchanges the indices ``first'' and ``second'' in the in 
the second fourfold integral of equation (\ref{pint}). Via interchanging 
these indices, we finally obtain $\int_{-\infty}^t{\rm d} 
t_1\int_{-\infty}^{t_1}{\rm d} t_2$ instead of initial $\int_{-\infty}^t{\rm d} 
t_1\int_{t_1}^{t}{\rm d} t_2$.
Taking into account these circumstances in the expression (\ref{pint}) for 
energy-momentum carried by electromagnetic field, we finally obtain
\begin{equation}\label{pnt}
p_{\rm em}^\alpha=\int\limits_{-\infty}^t{\rm d} 
t_1\int\limits_{-\infty}^{t_1}{\rm d} t_2
\int\limits_{0}^{k_1^0}{\rm d} R 
\int\limits_0^{2\pi}{\rm d}\varphi Jt^{0\alpha}
\end{equation}
where the new stress-energy tensor is symmetric in the pair of indices 1 and 
2:
\begin{eqnarray}\label{tfin}
2\pi t^{0\alpha}&=& 
2\pi\left(t^{0\alpha}_{12}+t^{0\alpha}_{21}\right)\\
&=&f_{(1)}^{0\lambda}f_{(2)\lambda}^\alpha +
f_{(2)}^{0\lambda}f_{(1)\lambda}^\alpha
 -\frac14\eta^{0\alpha}\left[
f_{(1)}^{\mu\nu}f^{(2)}_{\mu\nu} +f_{(2)}^{\mu\nu}f^{(1)}_{\mu\nu}
\right].\nonumber
\end{eqnarray}
We see that it is sufficient to consider the situation when $t_1\ge t_2$
pictured in figure \ref{bt}. The gray disc with radius $k_1^0=t-t_1$ is 
filled up by non-concentric circles with radii $R\in[0,k_1^0]$. Points in an 
$R$-circle are distinguished by angle $\varphi$.

To calculate the total flows (\ref{pnt}) of electromagnetic field energy and 
momentum  which flow across the plane $\Sigma_t$, we should integrate the 
Maxwell energy-momentum tensor density (\ref{tfin}) over angular variable 
$\varphi$, over radius $R$ and, finally, over time variables $t_1$ and 
$t_2$. Integration over $\varphi$ is a difficult undertaking. The difficulty 
resides mostly with norms  
$\|K_a\|^2=-\eta_{\alpha\beta}K_a^{\alpha}K_a^{\beta}$ of separation vectors 
$K_a=y-z_a$ which result in elliptic integrals. To avoid dealing with them 
we modify the coordinate transformation (\ref{ct}). We fix the parameter 
$\beta$ in such a way that the norm $\|K_1\|^2$ becomes proportional to the 
norm $\|K_2\|^2$:
\begin{equation} 
\|K_1\|^2=-\frac{\beta}{\alpha}\|K_2\|^2.
\end{equation} 
Keeping in mind the identity $\alpha+\beta=1$, we arrive at the quadratic 
algebraic equation on $\beta$ which does not contain the angle variable:
\begin{equation}\label{r2}
R^2=\alpha (k_1^0)^2+\beta (k_2^0)^2 -\alpha\beta{\bf q}^2.
\end{equation}
We choose the root which vanishes when $R=k_1^0$:
\begin{eqnarray}\label{beta}
\beta&=&\frac{1}{2{\bf q}^2}\left(
-(k_2^0)^2+(k_1^0)^2+{\bf q}^2+\sqrt{D}
\right)\\
D&=&\left[(k_2^0)^2-(k_1^0)^2-{\bf q}^2\right]^2 -
4{\bf q}^2\left[(k_1^0)^2-R^2\right].\nonumber
\end{eqnarray}
If ${\bf q}^2$ tends to zero while $t_1\ne t_2$, it becomes the unique 
root of the linear equation on $\beta$ originated from equation (\ref{r2}) 
with ${\bf q}^2=0$.

If $R=0$, the $R-$circle reduces to the point $A$ with coordinates 
$(z_1^i-\beta_0q^i)$, where $\beta_0=\left.\beta\right|_{R=0}$. If 
$R=k_1^0$ then $\beta=0$ and the circle is centred at $z^i(t_1)$
(see figure \ref{bt}). 

Differential chart of coordinate transformation (\ref{ct}) where $R(\beta)$ 
is given (implicitly) by equation (\ref{beta}) yields the Jacobian
\begin{equation}\label{Jc}
J=(1/2)\left[(k_2^0)^2-(k_1^0)^2-{\mathbf q}^2\right] +\beta {\mathbf 
q}^2-qR\cos\varphi.
\end{equation}

To calculate the energy-momentum (\ref{pnt}) carried by electromagnetic 
field, we should first perform the integration over angle. When facing 
this problem it is convenient to mark out $\varphi$-dependent terms 
in expressions under the integral sign. In the Maxwell energy-momentum 
tensor density (\ref{tfin}), we distinguish the second-order differential 
operator 
\begin{equation}\label{calTa}
{\hat{\cal T}}^a=
{\cal D}^a\frac{\partial^2 }{\partial t_1\partial t_2}+
{\cal B}^a\frac{\partial}{\partial t_1}
+{\cal C}^a\frac{\partial}{\partial t_2}+{\cal A}^a
\end{equation}
which has been labelled according to its dependence on combination of 
components of the separation vectors $K_1$ and $K_2$, or on the Jacobian 
(\ref{Jc}). The components of these vectors are involved in 
$\varphi$-dependent coefficients 
\begin{eqnarray}\label{ABC}
{\cal D}^a&=&\frac{1}{2\pi}\int_0^{2\pi}{\rm d}\varphi\frac{a}{r_1r_2},\qquad
{\cal B}^a=\frac{1}{2\pi}\int_0^{2\pi}{\rm d}\varphi\frac{ac_2}{r_1(r_2)^2}\\
{\cal C}^a&=&\frac{1}{2\pi}\int_0^{2\pi}{\rm d}\varphi\frac{ac_1}{(r_1)^2r_2},
\qquad
{\cal A}^a=\frac{1}{2\pi}\int_0^{2\pi}{\rm d}\varphi
\frac{ac_1c_2}{(r_1)^2(r_2)^2}
\nonumber
\end{eqnarray}
where the factor $a$ is replaced by $K_1^\alpha K_2^\beta, K_1^\alpha, 
K_2^\beta, J$ or $1$ for $\hat {\cal T}^{\alpha\beta}_{12}$, $\hat 
{\cal T}^\alpha_1$, $\hat {\cal T}^\beta_2$, $\hat {\cal T}^J$ or $\hat 
{\cal T}^0$, respectively.

To distinguish the partial derivatives in time variables, we rewrite the 
operator (\ref{calTa}) as the sum of the second-order differential operator
\begin{equation}\label{Pa}
{\hat\Pi}^a=\frac{\partial^2 }{\partial t_1\partial t_2}{\cal D}^a+
\frac{\partial}{\partial t_1}\left({\cal B}^a-
\frac{\partial{\cal D}^a}{\partial t_2}\right)
+\frac{\partial}{\partial t_2}\left({\cal C}^a-
\frac{\partial{\cal D}^a}{\partial t_1}\right)
\end{equation}
and the ``tail''
\begin{equation}\label{pa}
\pi^a=\frac{\partial^2 {\cal D}^a}{\partial t_1\partial t_2}-
\frac{\partial{\cal B}^a}{\partial t_1}-
\frac{\partial{\cal C}^a}{\partial t_2}+{\cal A}^a.
\end{equation}
For a smooth function $f(t_1,t_2)$ we have
\begin{equation}
{\hat{\cal T}}^a(f)={\hat\Pi}^a(f)+f\pi^a.
\end{equation}
Cumbersome calculations which is presented in \cite{YJMP} give the 
expressions
\begin{eqnarray}\label{cli}
\pi^0&=&0,\qquad \pi^J=0,\\
\pi_1^\alpha&=&v_1^\alpha\left({\cal B}^0-\frac{\partial{\cal 
D}^0}{\partial t_2}\right),\qquad
\pi_2^\beta=v_2^\beta\left({\cal C}^0-\frac{\partial{\cal D}^0}{\partial 
t_1}\right),
\nonumber\\
\pi_1^{\alpha J}&=&v_1^\alpha\left({\cal B}^J-\frac{\partial{\cal 
D}^J}{\partial t_2}\right),\qquad
\pi_2^{\beta J}=v_2^\beta\left({\cal C}^J-\frac{\partial{\cal 
D}^J}{\partial t_1}\right),
\nonumber\\
\pi_{12}^{\alpha\beta}&=&
v_1^\alpha\left({\cal B}_2^\beta-\frac{\partial {\cal D}_2^\beta}{\partial 
t_2}\right)
+v_2^\beta\left({\cal C}_1^\alpha-\frac{\partial {\cal 
D}_1^\alpha}{\partial t_1}\right)-v_1^\alpha v_2^\beta {\cal D}^0,
\nonumber
\end{eqnarray}
which allow us to rewrite the integral of $Jt^{\alpha\beta}$ over $\varphi$ 
in terms of differential operators ${\hat\Pi}^a$. So, the integral of energy 
density $t^{00}$ over the angular variable has the form
\begin{equation} \label{t00f}
\int_0^{2\pi}{\rm d}\varphi Jt^{00}=e^2\left[I{\hat\Pi
}^0(\kappa) - I'{\hat\Pi}^J(\mu)\right],
\end{equation}
where functions
\begin{eqnarray}\label{km}
\kappa&=&k_1^0k_2^0\frac{\partial^2\sigma}{\partial t_1\partial t_2} +
k_1^0\frac{\partial\sigma}{\partial t_1}+
k_2^0\frac{\partial\sigma}{\partial t_2}-
\frac12\frac{\partial\sigma}{\partial 
t_1}\frac{\partial\sigma}{\partial t_2}
+\sigma\mu,\\
\mu&=&\frac12\frac{\partial^2\sigma}{\partial t_1\partial t_2} +1 .
\nonumber
\end{eqnarray} 
World function $\sigma(t_1,t_2)$ of two timelike related points, 
$z(t_1)\in\zeta$ and $z(t_2)\in\zeta$, is equal to one-half of the square of 
vector $q^\mu=z^\mu(t_1)-z^\mu(t_2)$, taken with opposite sign:
\begin{equation}\label{sgm}
\sigma(\tau_1,\tau_2)=-\frac12(q\cdot q).
\end{equation} 
Symbols $I,I'$ denote $\beta$-dependent factors \begin{equation}\label{Ibar}
I=\frac{1}{\sqrt{-\beta\alpha}},\qquad 
I'=\sqrt{\frac{-\beta}{\alpha}}+
\sqrt{\frac{\alpha}{-\beta}}.
\end{equation}

The mixed space-time components of the stress-energy tensor ({\ref{tfin})
have the form
\begin{eqnarray} \label{t0if}
\int_0^{2\pi}{\rm d}\varphi Jt^{0i}&=&\frac{e^2}{2}I\left[
{\hat\Pi }_1^i\left(\frac{\partial\lambda_2}{\partial t_1}\right) +
{\hat\Pi }_2^i\left(\frac{\partial\lambda_1}{\partial t_2}\right)+
{\hat\Pi }^0\left(v_2^i\lambda_1+v_1^i\lambda_2\right)\right.\\
&-&\left.\frac{\partial}{\partial t_1}\left(v_2^i
\frac{\partial\lambda_1}{\partial t_2}{\cal D}^0
\right)-
\frac{\partial}{\partial t_2}\left(v_1^i
\frac{\partial\lambda_2}{\partial t_1}{\cal D}^0
\right)
\right]\nonumber\\
&-&\frac{e^2}{2}I'
{\hat\Pi}^J(v_1^i+v_2^i)\nonumber
\end{eqnarray} 
where
\begin{equation}\label{l1-2}
\lambda_1=k_1^0\frac{\partial\sigma}{\partial t_1}+\sigma,\qquad
\lambda_2=k_2^0\frac{\partial\sigma}{\partial t_2}+\sigma.
\end{equation}

We see that the integration of electromagnetic field's stress-energy 
tensor over $\varphi$ yields integrals being functions of the end points 
only. In the following subsection, we classify them and consider the problem 
of integration over the remaining variables.

\subsection{Integration over time variables and $\beta$}

Our purpose in this section is to develop the mathematical tools required in 
a surface integration of the energy-momentum tensor density in 
$2+1$ electrodynamics. Integration over angle variable results the 
combination of partial derivatives in time variables:
\begin{equation}\label{triple}
p_{\rm em}^\alpha(t)=\left.
\begin{array}{c}
\displaystyle 
e^2\int_{-\infty}^t {\rm d} t_1\int_{-\infty}^{t_1}{\rm d} t_2\\
\\[-1em]
\displaystyle
e^2\int_{-\infty}^t {\rm d} t_2\int_{t_2}^t {\rm d} t_1
\end{array}
\right\}
\left(
\frac{\partial^2 G^\alpha_{12}}{\partial t_1\partial t_2}+
\frac{\partial G^\alpha_1}{\partial t_1}+
\frac{\partial G^\alpha_2}{\partial t_2}
\right).
\end{equation}
Two double integrals over (proper) time variables (one about the other) 
describe integration over the domain $D_t=\{(t_1,t_2)\in{\mathbb 
R}^{\,2}: t_1\in ]-\infty,t],t_2\le t_1\}$.

By virtue of the equality
\begin{equation}\label{dG}
\int_{\beta_0}^0d\beta\frac{\partial G(\beta,t_1,t_2)}{\partial t_a}=
\frac{\partial}{\partial t_a}\left[\int_{\beta_0}^0d\beta
G(\beta,t_1,t_2)\right]+G(\beta_0,t_1,t_2)
\frac{\partial \beta_0(t_1,t_2)}{\partial t_a}
\end{equation}
the triple integral (\ref{triple}) can be rewritten as follows:
\begin{eqnarray}\label{pG}
p_{\rm em}^\alpha(t)&=&e^2\left[\lim_{k_1^0\to 0}
\int_{\beta_0}^0d\beta G_{12}^\alpha\right]_{t_2\to -\infty}^{t_2=t}
+e^2\int_{-\infty}^t{\rm d} t_2\lim_{k_1^0\to 0}\left[
\left.G_{12}^\alpha\right|_{\beta=\beta_0}
\frac{\partial \beta_0}{\partial t_2}\right]
\\
&-&e^2\int_{-\infty}^t{\rm d} t_2\lim_{\triangle t\to 0}
\int_{\beta_0}^0d\beta\left[\frac{\partial G_{12}^\alpha}{\partial t_2} 
+G_1^\alpha\right]_{k_1^0=k_2^0-\triangle t}
+e^2\int_{-\infty}^t{\rm d} t_2\lim_{k_1^0\to 0}\left[
\int_{\beta_0}^0d\beta G_1^\alpha\right]
\nonumber\\&+&e^2\int_{-\infty}^t{\rm d} t_1\lim_{\triangle t\to 0}\left[
\int_{\beta_0}^0d\beta G_2^\alpha\right]_{k_2^0=k_1^0+\triangle t 
}
-e^2\int_{-\infty}^t{\rm d} t_1\lim_{t_2\to -\infty}\left[
\int_{\beta_0}^0d\beta G_2^\alpha\right]
\nonumber\\&+&\left.
\begin{array}{c}
\displaystyle 
e^2\int_{-\infty}^t {\rm d} t_1\int_{-\infty}^{t_1}{\rm d} t_2\\
\\[-1em]
\displaystyle
e^2\int_{-\infty}^t {\rm d} t_2\int_{t_2}^t {\rm d} t_1
\end{array}
\right\}\left(
\left[\frac{\partial G_{12}^\alpha}{\partial t_2}+
G_1^\alpha\right]_{\beta=\beta_0}
\frac{\partial\beta_0}{\partial t_1} + 
\left.G_2^\alpha\right|_{\beta=\beta_0}\frac{\partial\beta_0}{\partial t_2}
\right).\nonumber
\end{eqnarray}
The functions under integral signs are as follows:
\begin{eqnarray}\label{G00}
G_{12}^0&=&I{\cal D}^0\kappa -
I'{\cal D}^J\mu,\\
G_1^0&=&-\sqrt{\frac{-\beta}{\alpha}}\kappa\frac{{\bf v}_1^2}{\|r_1\|^3}-
I'\mu\frac{\partial}{\partial\beta}
\left(\frac{\beta}{\|r_1\|}
\right),\nonumber\\
G_2^0&=&\sqrt{\frac{\alpha}{-\beta}}\kappa\frac{{\bf v}_2^2}{\|r_2\|^3}-
I'\mu\frac{\partial}{\partial\beta}
\left(\frac{\alpha}{\|r_2\|}
\right),\nonumber\\\label{Gi}
G_{12}^I&=&\frac{I}{2}\left[
\frac{\partial\lambda_1}{\partial t_2}{\cal D}_2^i
+\frac{\partial\lambda_2}{\partial t_1}{\cal D}_1^i
+\left(v_1^i\lambda_2+v_2^i\lambda_1\right){\cal D}^0\right]
- \frac{I'}{2}\left(v_1^i+v_2^i\right){\cal D}^J,\\
G_1^i&=&\frac{I}{2}\frac{\beta}{\|r_1\|^3}\left[
\frac{\partial\lambda_1}{\partial t_2}
\left(\alpha q^i{\bf v}_1^2+r_1^0v_1^i\right)+
\frac{\partial\lambda_2}{\partial t_1}
\left(-\beta q^i{\bf v}_1^2+r_1^0v_1^i\right)
+\left(v_1^i\lambda_2+v_2^i\lambda_1\right)
{\bf v}_1^2\right]\nonumber\\
&-&
\frac{I'}{2}\left(v_1^i+v_2^i\right)
\frac{\partial}{\partial\beta}
\left(\frac{\beta}{\|r_1\|}
\right),\nonumber\\
G_2^i&=&\frac{I}{2}\frac{\alpha}{\|r_2\|^3}\left[
\frac{\partial\lambda_1}{\partial t_2}
\left(\alpha q^i{\bf v}_2^2+r_2^0v_2^i\right)+
\frac{\partial\lambda_2}{\partial t_1}
\left(-\beta q^i{\bf v}_2^2+r_2^0v_2^i\right)
+\left(v_1^i\lambda_2+v_2^i\lambda_1\right)
{\bf v}_2^2\right]\nonumber\\&-&
\frac{I'}{2}\left(v_1^i+v_2^i\right)
\frac{\partial}{\partial\beta}
\left(\frac{\alpha}{\|r_2\|}
\right).\nonumber
\end{eqnarray}

All the terms involved in equation (\ref{pG}) possess specific small 
parameter. This circumstance allows us to expand the integrands into power 
series and perform the integration. 

{\bf $\mathbf 1^{\rm o}$. Integrals 
where $\mathbf t_1\to t$.} The lower limit $\beta_0$ tends to $0$ if 
$k_1^0=t-t_1$ vanishes. The upper limit is equal to zero too. Then the 
integral over parameter $\beta$ vanishes whenever an expression under 
integral sign is smooth. So, we must limit our computations to the singular 
terms only. These integrals do not contribute in the energy-momentum at all.

{\bf $\mathbf 2^{\rm o}$. Integrals where $\mathbf t_1=t_2$.} The small 
parameter is the positively valued difference $\triangle t=t_1-t_2$. The 
resulting terms belong to the bound part of energy-momentum. (To that which 
is permanently ``attached'' to the charge and is carried along with it.)

{\bf $\mathbf 3^{\rm o}$. Integrals where $\mathbf t_2\to-\infty$.} 
The lower limit $\beta_0$ tends to $0$ if $k_2^0=t-t_2$ increases extremely. 
Then the integral over parameter $\beta$ vanishes whenever an expression 
under integral sign is smooth. So, we must limit our computations to the 
singular terms only. The resulting terms belong to the bound electromagnetic 
``cloud'' which can not be separated from the charged particle.

{\bf $\mathbf 4^{\rm o}$. Integrals at point where $\mathbf 
\beta=\beta_0$.} In this case, the radius of the smallest circle pictured 
in figure \ref{bt} vanishes and it reduces to the point $A$. The 
contribution in $p_{\rm em}^\alpha$ is given by the last line of 
equation (\ref{pG}). It can be rewritten as the combination of partial 
derivatives in time variables and non-derivative ``tail''. After integration 
over $t_1$ or $t_2$, the derivatives are coupled with bound terms; the sum 
is absorbed by three-momentum of ``bare'' particle within the 
renormalization procedure. The ``tail'' contains radiative terms which 
detach themselves from the charge and lead an independent existence. 

Summing up all the contributions $1^{\rm o}-4^{\rm o}$, we finally obtain
\begin{eqnarray}\label{p0em}
p^0_{\rm em}(t)&=&e^2\left.
\frac{1+1/2\sqrt{1-{\bf v}_1^2}}{1+\sqrt{1-{\bf v}_1^2}}\frac{1}{\sqrt{1-{\bf v}_1^2}}
\right|_{t_1\to -\infty}^{t_1=t}+
\frac{e^2}{2}\int_{-\infty}^t{\rm d} t_2\frac{1}{\sqrt{2\sigma(t,t_2)}}
\\&+&e^2\int_{-\infty}^t {\rm d} t_1\int_{-\infty}^{t_1}{\rm d} t_2
\left[
-\frac{(v_1\cdot v_2)q^0}{[2\sigma(t_1,t_2)]^{3/2}}+
\frac12\frac{(v_1\cdot q)}{[2\sigma(t_1,t_2)]^{3/2}}+
\frac12\frac{(v_2\cdot q)}{[2\sigma(t_1,t_2)]^{3/2}}
\right]\nonumber\\\label{piem}
p^i_{\rm em}(t)&=&e^2\left.
\frac{1+1/2\sqrt{1-{\bf v}_1^2}}{1+\sqrt{1-{\bf 
v}_1^2}}\frac{v_1^i}{\sqrt{1-{\bf v}_1^2}}
\right|_{t_1\to -\infty}^{t_1=t}+
\frac{e^2}{2}\int_{-\infty}^t{\rm d} t_2\frac{v_2^i}{\sqrt{2\sigma(t,t_2)}}
\\&+&e^2\int_{-\infty}^t {\rm d} t_1\int_{-\infty}^{t_1}{\rm d} t_2
\left[
-\frac{(v_1\cdot v_2)q^i}{[2\sigma(t_1,t_2)]^{3/2}}+
\frac12\frac{(v_1\cdot q)v_2^i}{[2\sigma(t_1,t_2)]^{3/2}}+
\frac12\frac{(v_2\cdot q)v_1^i}{[2\sigma(t_1,t_2)]^{3/2}}
\right]\nonumber
\end{eqnarray}
where $\sigma$ denotes the two-point function (\ref{sgm}). The finite terms 
which depend on the end points only are non-covariant. They express the 
``deformation'' of electromagnetic ``cloud'' due to the choice of the
coordinate-dependent hole around the particle in the integration surface 
$\Sigma_t$. We neglect these structureless terms. The single integrals 
describe covariant singular part of energy-momentum carried by 
electromagnetic field. They arise from the following sum of ``three-point 
functions'' which depend on particle's position and velocity referred to the 
instants $t_1$ and $t_2$ before observation instant $t$ as well as on $t$ 
itself:
\begin{eqnarray}\label{sing}
\frac{e^2}{2}\int_{-\infty}^tdt_2\frac{v_2^\mu}{\sqrt{2\sigma(t,t_2)}}&=&
\frac{e^2}{2}\int_{-\infty}^t dt_2 
\left.
\frac{v_2^\mu}{\sqrt{(2t-t_1-t_2)^2-{\mathbf q}^2}}
\right|_{t_1=t_2}^{t_1=t}\\
&+&\frac{e^2}{2}\int_{-\infty}^t dt_1
\left.
\frac{v_1^\mu}{\sqrt{(2t-t_1-t_2)^2-{\mathbf q}^2}}
\right|_{t_2\to -\infty}^{t_2=t_1}.\nonumber
\end{eqnarray}
(It is worth noting that the denominator evaluated at the remote past when
$t_2\to -\infty$ vanishes even if $t_1\to -\infty$ too.) Two-point functions 
in between the square brackets of equations (\ref{p0em}) and (\ref{piem}) 
determine the radiation reaction in $2+1$ electrodynamics.


\begin{thebibliography}{99} 
\bibitem{Gl}
Gal'tsov D V 2002 {\it Phys. Rev.} D {\bf 66} 025016

\bibitem{KLS}
Kazinski P O, Lyakhovich S L and Sharapov A A 2002 {\it Phys. Rev.} D {\bf 66} 025017

\bibitem{Dir}
P.A.M.Dirac, Proc.Roy.Soc.A {\bf 167}, 148 (1938).

\bibitem{Rohr}
F.Rohrlich, {\it Classical Charged Particles}, 2nd ed. (Redwood City, CA:
Addison-Wesley, 1990).

\bibitem{TVW}
Teitelboim C, Villarroel D and van Weert C C 1980 {\it Riv.Nuovo Cimento} 
{\bf 3} 9

\bibitem{PsPr}
Poisson E 1999  {\it An introduction to the Lorentz-Dirac equation}  
arXiv:gr-qc/9912045 

\bibitem{WB}
DeWitt B S and Brehme R W 1960 {\it Ann.Phys. (N.Y.)} {\bf 9} 220

\bibitem{Hb}
Hobbs J M 1968 {\it Ann.Phys. (N.Y.)} {\bf 47} 141

\bibitem{DW}
Detweiler S and Whiting B F 2003 {\it Phys. Rev.} D {\bf 67} 024025

\bibitem{Teit}
C.Teitelboim, Phys.Rev.D {\bf 1}, 1572 (1970).

\bibitem{LV}
C.A.L\'opez and D.Villarroel,  Phys.Rev.D {\bf 11}. 2724 (1975).

\bibitem{Yar03}
Yu.Yaremko, J. Phys.A:Math.Gen. {\bf 36}, 5149 (2003).

\bibitem{Yar04}
Yaremko Yu 2004 {\it J.Phys.A: Math. Gen.} {\bf 37} 1079

\bibitem{Kos}
Kosyakov B P 1999 {\it Teor. Mat. Fiz.} {\bf 119} 119 (in Russian)

Kosyakov B P 1999 {\it Theor. Math. Phys.} {\bf 119} 493 (Engl.Transl.);
arXiv:hep-th/0208035

\bibitem{QW}
Quinn T C and Wald R M 1999 {\it Phys. Rev.} D {\bf 60} 064009

\bibitem{MST}
Mino Y, Sasaki M and Tanaka T 1997 {\it Phys. Rev.} D {\bf 55} 3457

\bibitem{Pois}
Poisson E 2004 {\it Living Rev. Relativity} {\bf 7} Irr-2004-6;
arXiv:gr-qc/0306052 

\bibitem{YJMP}
Yaremko Yu 2007 {\it J.Math.Phys.} {\bf 48} 092901

\bibitem{Br}
Burko L M 2002 {\it Class. Quant. Grav.} {\bf 19} 3745

\bibitem{AHNS}
Ambegaokar V, Halperin B I, Nelson D.R, Siggia E D 1980 {\it Phys. Rev.} B 
{\bf 21} 1806

\bibitem{FL}
Fisher M P A and Lee D H 1989 {\it Phys. Rev.} B {\bf 39} 2756

\bibitem{Z}
Zhang S-C 2002 {\it To see a world in a grain of sand} 
arXiv:hep-th/0210162 

\end{thebibliography}
\end{document}